# Electron-phonon coupling and phonon dynamics in single-layer NbSe$_2$ on graphene: the role of moiré phonons

Amjad Al Taleb[1], Wen Wan[2], Giorgio Benedek[2,3],
Miguel M. Ugeda[2,4,5] and Daniel Farías[1,6,7,*]

[1]*Departamento de Física de la Materia Condensada, Universidad Autónoma de Madrid, 28049 Madrid, Spain*
[2] *Donostia International Physics Center, Paseo Manuel de Lardizábal 4, 20018 San Sebastián, Spain.*
[3]*Dipartimento di Scienza dei Materiali, Università di Milano-Bicocca, 20125 Milano, Italy*
[4]*Ikerbasque, Basque Foundation for Science, 48013 Bilbao, Spain.*
[5]*Centro de Física de Materiales, Paseo Manuel de Lardizábal 5, 20018 San Sebastián, Spain.*
[6]*Instituto Nicolás Cabrera, Universidad Autónoma de Madrid, 28049 Madrid, Spain*
[7]*Condensed Matter Physics Center (IFIMAC), Universidad Autónoma de Madrid, 28049 Madrid, Spain*

## Abstract

*The interplay between substrate interactions and electron-phonon coupling in two-dimensional (2D) materials presents a significant challenge in understanding and controlling their electronic properties. Here, we present a comparative study of the structural characteristics, phonon dynamics, and electron-phonon interactions in bulk and monolayer NbSe$_2$ on epitaxial bilayer graphene (BLG) using helium atom scattering (HAS). High-resolution helium diffraction reveals a (9x9)0º superstructure within the NbSe$_2$ monolayer, commensurate with the BLG lattice, while out-of-plane HAS diffraction spectra indicate a low-corrugated (3√3×3√3)30º substructure. By monitoring the thermal attenuation of the specular peak across a temperature range of 100 K to 300 K, we determined the electron-phonon coupling constant ($\lambda_{HAS}$) as 0.76 for bulk 2H-NbSe$_2$. In contrast, the NbSe$_2$ monolayer on graphene exhibits a reduced $\lambda_{HAS}$ of 0.55, corresponding to a superconducting critical temperature ($T_C$) of 1.56 K according to the MacMillan formula, consistent with transport measurement findings. Inelastic HAS data provide, besides a set of dispersion curves of acoustic and lower optical phonons, a soft,*

*dispersionless branch of phonons at 1.7 meV, attributed to the interface localized defects distributed with the superstructure period, and thus termed moiré phonons. Our data show that moiré phonons contribute significantly to the electron-phonon coupling in monolayer NbSe$_2$. These results highlight the crucial role of the BLG on the electron-phonon coupling in monolayer NbSe$_2$, attributed to enhanced charge transfer effects, providing valuable insights into substrate-dependent electronic interactions in 2D superconductors.*

**I. Introduction**

Layered transition-metal chalcogenides (TMD) are well-suited systems to study the effects of dimensionality and thickness on collective electronic phases like superconductivity and charge density waves (CDW). These are generally manifestations of electron-phonon interactions, leading to electron-electron (hole-hole) pairing and electron-hole ordering, respectively. Bulk 2H-NbSe$_2$ is especially interesting due to the coexistence of superconducting and CDW phases: a CDW with a (3x3) periodicity is known to set in below 33 K [1], while superconductivity sets in at $T_c$ = 7.2 K [2]. In the monolayer NbSe$_2$ limit, different results have been reported depending on the supporting substrate. On epitaxial bilayer graphene (BLG) on SiC(0001), $T_C$ is depressed to 1.5 K whereas the critical temperature for the appearance of the CDW remains unchanged (33 K) [3]. However, when the monolayer is hold on sapphire, superconductivity was observed to set in at $T_c$ = 3 K, while the CDW critical temperature increases to 145 K [4]. These results were interpreted in terms of an increase in the electron-phonon coupling in single-layer NbSe$_2$/sapphire. Regarding charge transfer, there is little difference between sapphire (work function ~4.5 eV [5]) and BLG/SiC (work function 4.30 eV [6]), when compared to NbSe$_2$ work function of 5.9 eV [7]. Thus, these different behaviors in the presence of similar charge transfer raise the question about the possible role of BLG underneath in both superconductivity and CDW and more specifically, on the strength of the electron-phonon coupling in single-layer NbSe$_2$.

In this work, we report measurements of the electron-phonon coupling constant (mass-enhancement factor) $\lambda_{HAS}$ in single-layer NbSe$_2$/BLG/SiC(0001) by high-resolution helium atom scattering (HAS) spectroscopy [8], which enabled us to extract information about the surface structure on a long-period (nm) scale and related low-energy (meV) dynamics. Our He-diffraction results show that single-layer NbSe$_2$ on BLG develops a commensurate superstructure, where a (9x9) NbSe$_2$ monolayer supercell matches with a small tensile strain a (13x13) BLG supercell. A comparatively intense, dispersionless low-energy (1.7 meV) phonon branch is also observed. This effect induced by the superstructure on phonon dynamics is analogous to what observed in the electronic structure, like the occurrence of ultra-flat electronic bands, e.g., in twisted bilayers of WSe$_2$ [9] and of graphene [10,11], ultimately responsible for the observed superconductivity in in these heterostructures.

A recent quantum-theoretical approach showed how the thermal attenuation of the HAS specular peak from metal surfaces, described by the Debye-Waller (DW) exponent, can directly provide the electron-phonon coupling constant (mass-enhancement factor $\lambda$, here denoted $\lambda_{HAS}$ when obtained with this method) [12] This method to derive $\lambda_{HAS}$ has been recently extended to other classes of conducting surfaces, like those of layered TMDs [13,14] and supported graphene monolayers [15], and is here applied to the present NbSe$_2$ heterostructures to elucidate the role of the low-energy phonon branch in the electron-phonon coupling. In general, HAS spectroscopy should be relevant for the characterization of twisted layered heterostructures and their functional properties in the new area of *twistronics* [16].

## II. Results and discussion

### II.a) Diffraction

Figure 1(a) shows the HAS angular distribution measured for single-layer NbSe$_2$/BLG/SiC(0001) (red curve) along the (1000) direction, and is compared to that for a 2-3 layers of NbSe$_2$/BLG/SiC(0001) (green curve) and for the bulk NbSe$_2$(0001) surface (black curve), all recorded at $T = 100$ K. The spectrum for bulk NbSe$_2$ shows numerous intense diffraction peaks, indicating a highly corrugated surface, similar to those typically found on semiconductor surfaces. A comparable diffraction spectrum is observed for the few-layer sample, while the monolayer NbSe$_2$ exhibits a significantly different pattern. In this case, the diffraction intensities are much weaker than the specular scattering, suggesting a metallic surface for the monolayer due to significant charge transfer from the substrate. This is consistent with the comparatively large difference between the work function of bulk NbSe$_2$ (5.9 eV) [17] and that of few-layer NbSe$_2$/BLG (4.30 eV)[18].

The charge trasfer is also responsible of a small shift of the diffraction peaks of the NbSe$_2$ monolayer as compared to those of the bulk and few-layer samples. For the $(0\bar{1})$ diffraction the latter occur at $\theta_f = 39.5°$, which gives an in-plane lattice constant $a_N = 3.44$ Å, in perfect agreement with the value measured in NbSe$_2$ multilayer flakes [19], while the $(0\bar{1})$ peak maximum for the monolayer is shifted to 39.9°. This corresponds to $a_N = 3.52$ Å, i.e., to a ~2.3% expansion of the in-plane monolayer lattice constant. The electronic charge transfer leaves behind a positive potential in the substrate and, therefore, a component in the surface charge density distribution reflecting the substrate periodicity. Actually, the ×4 magnification of the diffraction in the monolayer (Fig. 1(b)) hints to a maximum around $\theta_f = 33°$, where the BLG $(0\bar{1})$ diffraction peak is expected to occur for the known graphene lattice parameter of 2.46 Å [20,21].

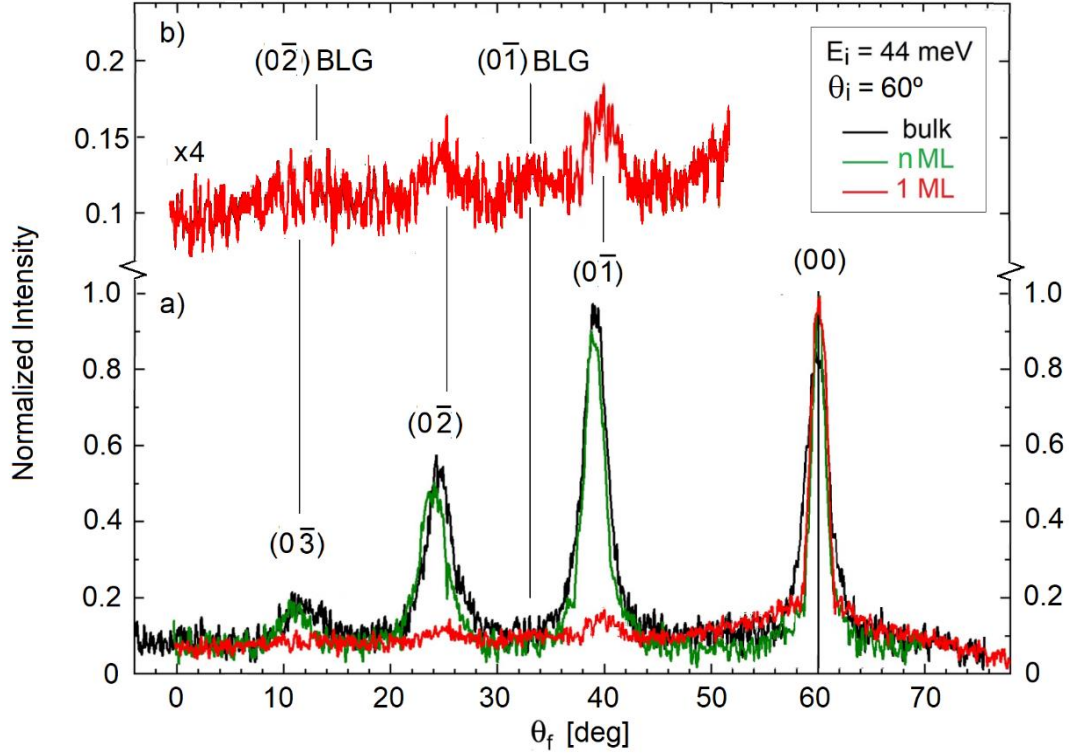

**Figure 1.** (a) Angular distribution of HAS along the (10) direction ($\Gamma$M) of the surfaces of bulk NbSe$_2$(0001) (black curve), few-layer NbSe$_2$/BLG/SiC(0001) (green curve) and single-layer NbSe$_2$/BLG/SiC(0001) (red curve). The angle of incidence is 60° and the incident energy of the He beam is $E_i = 44$ meV. (b) A factor-4 expansion of the diffraction spectrum for the monolayer NbSe$_2$, where the BLG $(0\bar{1})$ diffraction peak is also visible, as an effect of the charge transfer.

The observation of the diffraction peaks of both the monolayer NbSe$_2$ and the BLG substrate induced by a large charge transfer is confirmed by the high-resolution HAS diffraction pattern displayed in Fig. 2. The $(0\bar{1})$ BLG diffraction peak at $\varphi = -22.1°$ is about one half in size as that of single-layer NbSe$_2$ at $\varphi = -15.2°$ and comparable in sharpness, and corresponds to the in-plane lattice constants $a_{BLG} = 2.47 \pm 0.02$ Å and $a_N = 3.55 \pm 0.02$ Å. This value of $a_N$ confirms the in-plane expansion (here 3.2 ±0.6%) of the NbSe$_2$ monolayer with respect to the bulk induced by the charge transfer.

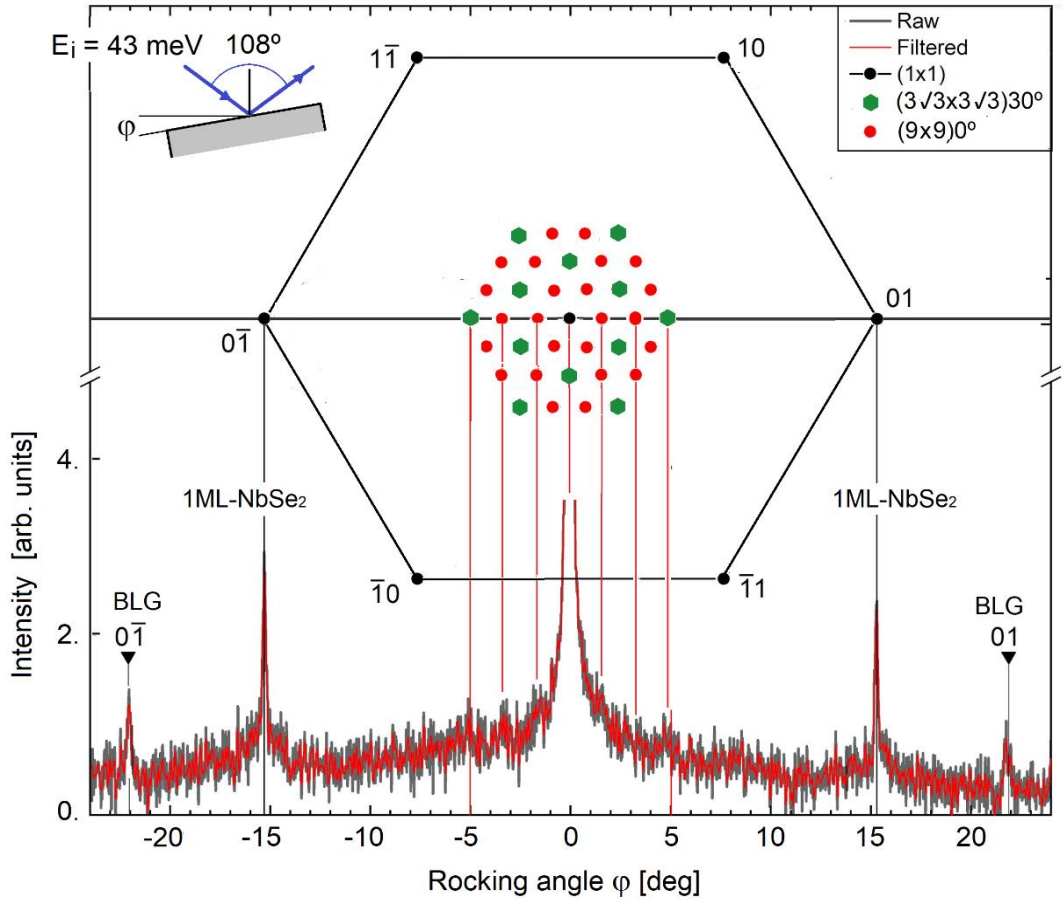

**Figure 2**. High-resolution HAS angular distribution measured as a function of the rocking angle (top-left inset) along the (10) direction (ΓM) of single-layer NbSe$_2$/BLG/SiC(0001) at T = 100 K, with a fixed scattering angle of 108° and incident energy of 23 meV. The upper part of the figure shows the corresponding diffraction spots in the reciprocal space

of single-layer NbSe$_2$ (black dots) and those near the specular peak of the single-layer NbSe$_2$/BLG superstructures indicated in the top-right inset and discussed in the text.

Since the diffraction components of both the single-layer NbSe$_2$ and BLG are observed in Fig. 2, the small features at $\varphi \cong$ -5º, -3.3º, -1.7º and 5º near the specular peak and discernible above the noise (especially in the filtered data, red line), is assigned to a commensurate superstructure with a unit cell aligned with those of the two component lattices. The lattice constant ratio $a_N/a_{BLG}$ = 1.44 is very close to 13/9 = $1.\overline{4}$, suggesting that the hexagonal superlattice unit cell is fromed by the coincidence of a (9×9)0º cell of the NbSe$_2$ monolayer with a (13×13)0º cell of BLG. This is illustrated by Fig. 3(a), which shows the interface Se ions (green atoms) positioned over the graphene (grey atoms), with the superlattice unit cell edges marked in red. The corresponding theoretical diffraction spots around the origin (specular peak) in the reciprocal plane are shown in the upper part of Fig. 2 (red and green dots), with some spots closely corresponding to the small features described above.

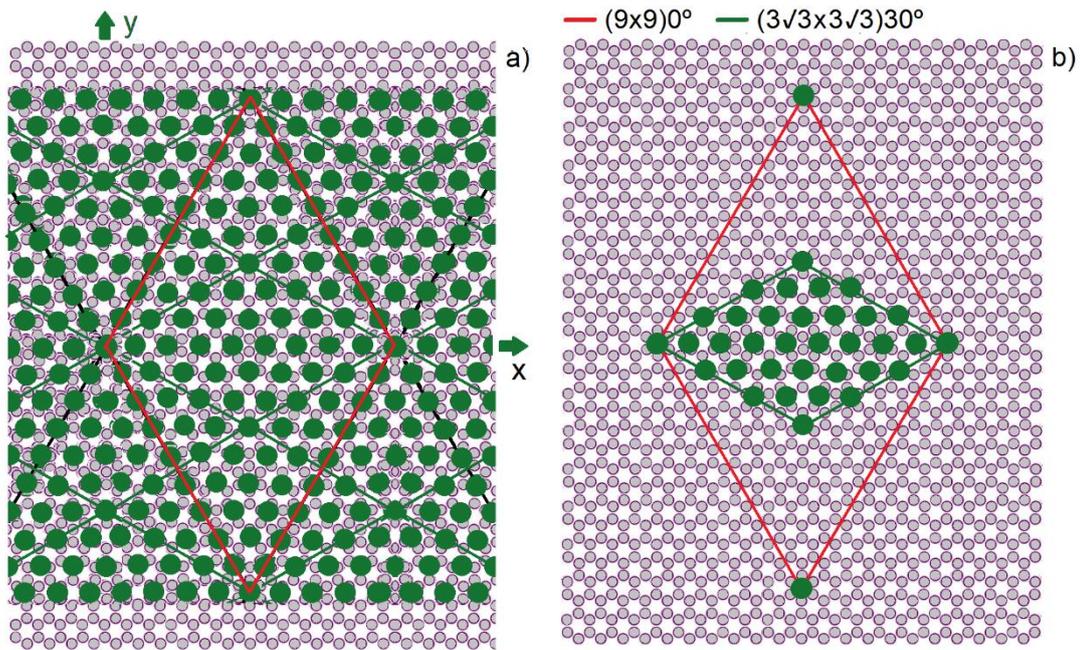

**Figure 3**. (a) Structure of the single-layer NbSe2/BLG interface showing the Se ions (green full circles) on graphene (grey C atoms). The red lines show the unit cell of the single-layer NbSe2 (9×9)0º supercell matching the (13×13)0º supercell of BLG. (b) The reduced (3√3×3√3)30º unit cell of single-layer NbSe2 (green contour line) in case of equivalence between the BLG high-symmetry hexagonal sites (left and right corners of the reduced cell) and the trigonal symmetry sites (top and bottom corners), approximately produced by the addition of the second BLG graphene layer (not shown).

The presence of a second graphene layer underneath causes the Se ions at the hexagonal symmetry sites (e.g., the corners of the (9×9)0º unit cell in Fig. 3(a)) to be almost equivalent, in terms of adsorption energy, to those at the trigonal symmetry sites (e.g., the top and bottom atoms of the reduced (3√3×3√3)30º unit cell, as shown in Fig. 3(b)). If these sites were strictly equivalent, the actual (3√3×3√3)30º supercell would be three times smaller, and only the diffraction peaks corresponding to the green spots in

Fig. 2 would appear. However, even with approximate equivalence, the green spots—especially the first six, which form a 30º rotated hexagon in Fig. 2—are reinforced.

Out-of-plane measurements made using the HAS confirm this effect. The HAS intensity, mapped as a function of the final angle $\theta_f$ and out-of-plane azimuth angle $\varphi_f$ for a fixed incident energy $E_i = 56$ meV and incident angle $\theta_i = 60º$ (see Fig. 4(b); kinematics explained in Fig. 4(a)), shows maxima at the expected positions for the $(3\sqrt{3}\times3\sqrt{3})30º$ supercell (green hexagons). Note that the hexagonal pattern is slightly rotated clockwise relative to the calculated maxima positions, due to a minor azimuthal misalignment of the NbSe$_2$ lattice with respect to the sagittal plane $\varphi_i = 0º$ (see Fig. SI-1 for details). However, this small instrumental misalignment does not undermine the strong evidence supporting the $(3\sqrt{3}\times3\sqrt{3})30º$ supercell as the approximate commensurate structure for single-layer NbSe$_2$/BLG/SiC(0001).

It is worth mentioning that essentially the same diffraction spectra for the NbSe$_2$ monolayer, including the out-of-plane map demonstrating the $(3\sqrt{3}\times3\sqrt{3})30º$ superlattice, have been measured at $T = 40$ K (see Fig. SI-1), which indicates that there is no evidence for a (3x3) CDW pattern above 40 K. This is in agreement with previous STM results on the same system, which showed that a CDW sets in below 35 K [3], and in clear contrast with previous Raman measurements using samples exposed to ambient conditions where $T_C = 145$ K [4].

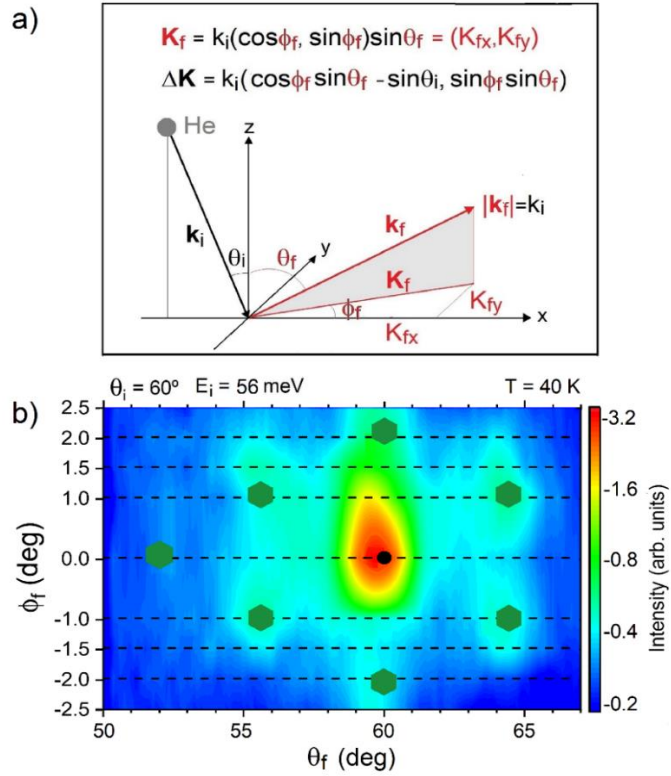

**Figure 4.** (a) Geometry and kinematics of out-of-plane HAS diffraction relating the components of the final parallel wavevector $\mathbf{K}_f$ and parallel wavevector transfer $\Delta\mathbf{K}$ to the incident wavevector $\mathbf{k}_i$ (or incident energy through $k_i[\text{Å}^{-1}] = 1.383\sqrt{E_i[\text{meV}]}$ for $^4$He atoms [8]) and scattering geometry. (b) The out-of-plane HAS diffraction map of single-layer $NbSe_2$/BLG/SiC(0001) at $T = 40$ K, measured as a function of the final and azimuthal angles defined in (a) at a fixed incident angle of 60º and energy of 52 meV. The 30º-rotated hexagonal pattern reflects the theoretical positions of the diffraction peaks (green hexagonal dots) for the $(3\sqrt{3}\times 3\sqrt{3})30°$ superlattice.

**II.b) The Debye-Waller exponent**

Figure 5 shows the temperature dependence of the Debye-Waller exponent, as derived from the ratio of He specular intensity at the surface temperature $T$, $I(T) \equiv I$, to that at the lowest measured temperature $T_0$, $I(T_0) \equiv I_0$, for single-layer $NbSe_2$ compared to the data for bulk NbSe$_2$ and the few-layer NbSe$_2$. The data for the NbSe$_2$ overlayers are reported for two different He beam incident energies and angles.

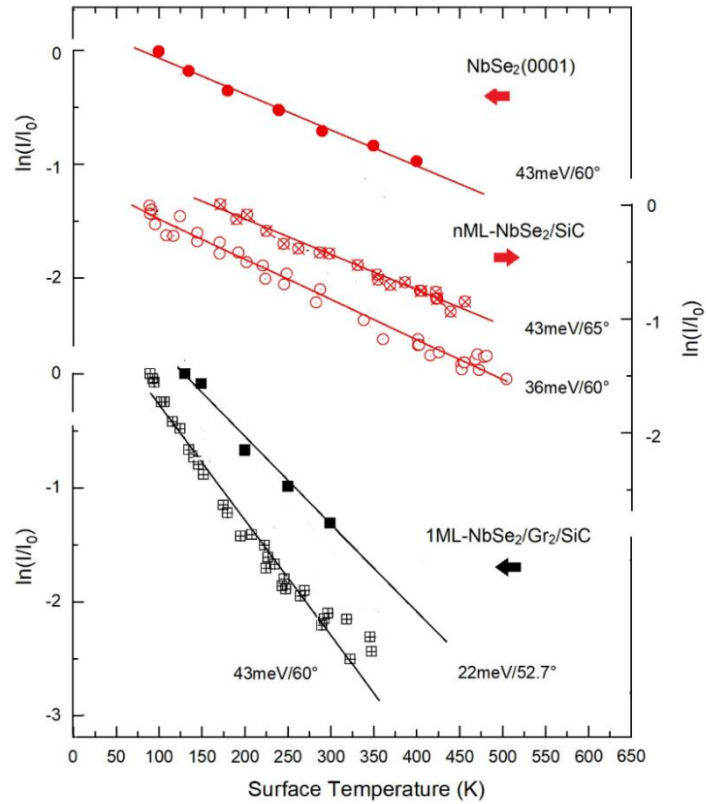

**Figure 5.** Thermal attenuation of He-specular intensity measured along the ΓM azimuth from bulk NbSe$_2$ (top), few-layer NbSe$_2$/BLG/SiC(0001) (middle), and for single-layer NbSe$_2$/BLG/SiC(0001) (bottom). Note the different ordinate scales for the three sets of data. The corresponding incident energy and incident angles are given for each set of data. The straight lines correspond to the best linear fits according to Eq. (2).

While the exponential attenuation of the specular intensity is about the same for the semi-infinite crystal and for the 2-3 layers, a much steeper attenuation is observed for the NbSe$_2$ monolayer on top of the graphene bilayer. This is mainly due to the graphene substrate, which provides several intervalley channels, as discussed in previous HAS studies on the electron-phonon interaction of graphene on various substrates [15]. In the next subsection, the corresponding values of the electron-phonon coupling constant are derived for the present set of data, based on the information from phonon dispersion curves derived from HAS time-of-flight (TOF) data. While HAS data for bulk NbSe$_2$ phonons are already available [22] and assumed to be approximately valid for multilayers, the HAS phonon data for the monolayer NbSe$_2$ are presented in the following subsection.

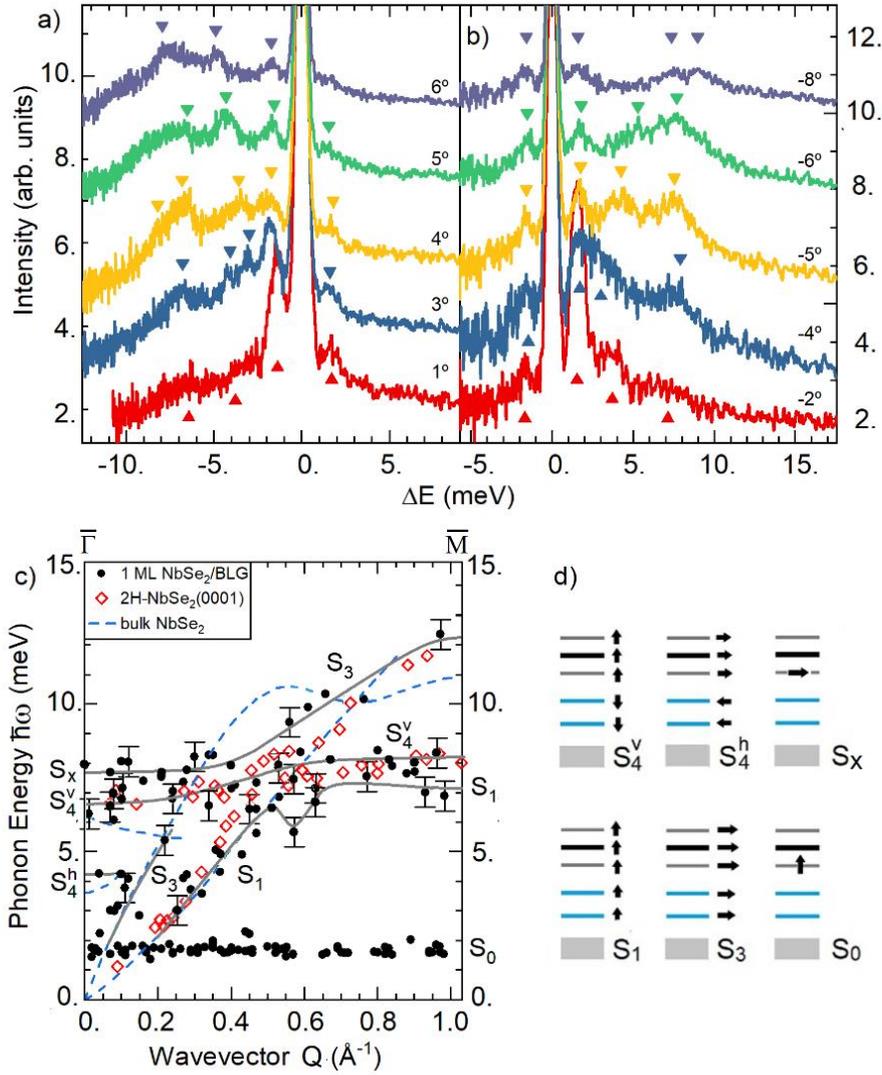

**Figure 6.** (a,b) Selected time-of-flight spectra measured with HAS at constant scattering angle $\theta_i + \theta_f = 108°$, incident energy $E_i = 26$ meV, along the ΓM direction of single-layer NbSe$_2$/BLG at 100 K; each spectrum is labelled by the rocking angle $\varphi$ of the surface plane, so that $\theta_i = 54° + \varphi$ and $\theta_f = 54° - \varphi$. (c) The corresponding phonon dispersion

curves (full black circles, with full grey lines as eye guidelines) are compared to those measured with HAS for the semi-infinite NbSe$_2$(0001) surface at room temperature (red open symbol [22], and to the bulk NbSe$_2$ dispersion curves (blue broken lines) [23]. Besides the surface branches S$_1$, S$_3$, S$_4^v$ and S$_4^h$, a flat soft branch S$_0$ is observed in single-layer NbSe$_2$/BLG at 1.7 meV, attributed to interface localized (moiré) modes with vertical (z) polarization. The horizontal (x-polarized) counterpart is associated with the S$_x$ branch at 8 meV, which converts via avoided crossing to the S$_3$ branch at about ΓM/2. The displacement patterns of the atomic planes for the different branches are schematically shown in (d), with the BLG illustrated by blue segments, the NbSe$_2$ three atomic layers by grey and black segments, and the SiC(0001) substrate, assumed to be rigid, by grey boxes.

## II.c) Phonons

Two of the several sets of HAS TOF spectra of 1ML-NbSe$_2$/BLG/SiC(0001), measured at a surface temperature of 100 K along the ΓM direction, are shown in Figs. 6(a,b) as functions of the energy transfer $\Delta E_i$. The inelastic peaks in the TOF spectra, like those marked by small triangles in Figs. 6(a,b) for phonon creation ($\Delta E_i < 0$) and annihilation ($\Delta E_i > 0$) processes, provide the phonon energy $\hbar\omega = |\Delta E_i|$ as a function of the parallel wavevector $Q = |\Delta K|$ (full black circles in Fig. 6c). A possible set of dispersion curves is represented by the eye-guidelines (gray full lines) drawn in Fig. 6(c). They are compared to the data for the 2H-NbSe$_2$(0001) surface (red open symbols) [22] and to the bulk NbSe$_2$ dispersion curves (blue broken lines) [23].

In the long-wave limit the surface phonon branches in the acoustic region of 1ML-NbSe$_2$/BLG labelled as S$_1$, S$_3$, S$_4^v$ and S$_4^h$ correspond to the Rayleigh wave, the longitudinal resonance, the shear-vertical and longitudinal oscillations of the NbSe$_2$ monolayer against the BLG (Fig. 6(d)). In 2H-NbSe$_2$(0001) the surface unit cell includes two NbSe$_2$ monolayers, while in our system the BLG plays, to some extent, the role of the second monolayer. This may alone explain the softening of the S$_1$ branch and the stiffening of S$_3$, both in the second half of ΓM, with respect to those of bulk 2H-NbSe$_2$(0001) (red open symbols), although the stiffening of S$_3$ may also be related to the avoided crossing with S$_x$. Note that a possible Kohn anomaly occurs in the S$_1$ branch near 0.6 Å$^{-1}$, more pronounced than that found with HAS at room temperature in bulk NbSe$_2$(0001) [22], and showing a shift to a smaller $Q$ with respect to that reported with neutron scattering in bulk at $^2/_3$ΓM (blue broken line) [23]. This is probably related to the charge transfer and the consequent shift of the Fermi wavevector. A similar shift of the anomaly has been observed with HAS in 2H-TaSe$_2$(0001) for the Rayleigh mode with respect to the anomaly in the bulk acoustic mode [24].

The low-energy flat phonon branch S$_0$, displaying a strong intensity at small $Q$ (Figs. 6(a.b)) may be associated to soft shear-vertical vibrations localized at a superlattice periodic array of atoms, for example the interface Se ions which accommodate at the high symmetry hexagonal (and possibly trigonal) sites of graphene (corner atoms in Fig. 3(b)) with strongly weakened local shear force constants. Note that this branch is reminiscent of the low energy dispersionless modes observed on Gr/Ir(111) and Gr/Ru(0001), where graphene builds a moiré pattern [25,26]. Due to this analogy and the present interpretation of the flat branch as closely associated with the interface superstructure, in the following discussion we shall term these localized excitations as *moiré* phonons. On the other hand, such localized shear vertical vibrations should have their longitudinal companion, which is here associated with the S$_x$ additional branch, also localized above the maximum of the acoustic band in the region around the zone center [27]. Clearly, only a surface dynamics first-principle calculation for this system would allow for a reliable assignment of the observed additional moiré branches $S_0$ and $S_x$.

**II.d) Electron-phonon interaction**

The HAS data for $\ln(I/I_0)$, shown in Fig. 5 for the surfaces studied, provide the HAS Debye-Waller (DW) exponent $-2W(T)$, up to a constant that depends on how $I_0$ is defined. Therefore, the slope of $\ln(I/I_0)$ as a function of temperature directly represents the slope of $-2W(T)$, regardless of the specific definition of $I_0$. This slope can be used to determine the electron-phonon coupling constant, $\lambda_{HAS}$ (also known as the mass-enhancement factor), for bulk $NbSe_2$ and the few-layers. This approach is based on the method valid in the high-temperature limit [12].

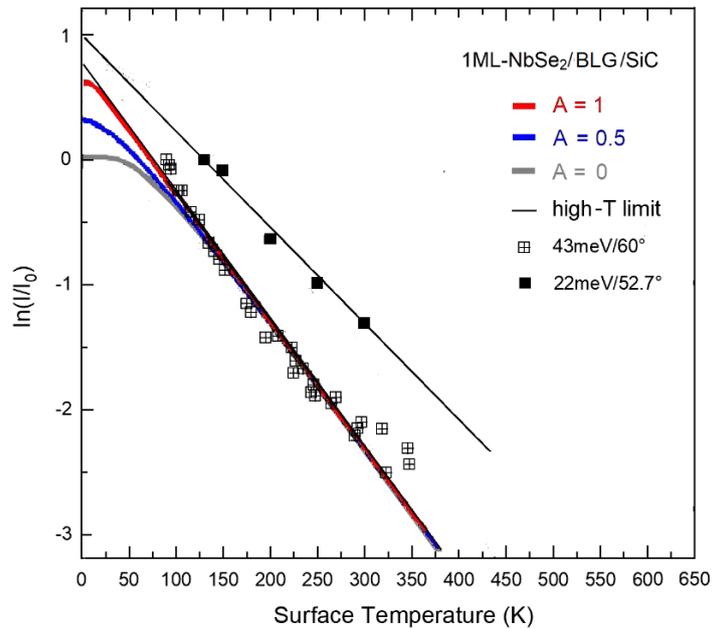

**Figure 7.** Fits of the Debye-Waller data for HAS incident energy $E_i = 43$ meV and angle $\theta_i = 60°$ using Eq. (1). The red curve represents a fit including only the moiré phonons contribution ($A = 1$), the blue curve represents an equal share of the moiré and the NbSe$_2$ phonons ($A = 0.5$) and the grey curve includes only the NbSe$_2$ ML phonons ($A = 0$). The few data for $E_i = 22$ meV and $\theta_i = 52.7°$, all above 125 K, are well fitted by the high-$T$ straight line, independently of $A$.

For single-layer NbSe$_2$/BLG/SiC, at least two different contributions to $\lambda_{HAS}$ are expected from phonons from the NbSe$_2$ monolayer and moiré phonons from the interface. The high-energy optical phonons of the BLG can be neglected, however, due to their small population in the present temperature range, and their depth beneath the first surface atomic layer. In this approximation, the expression of the DW exponent receiving contributions from two distinct phonon spectral regions can be written as in Ref. [28]

$$2W(\mathbf{k}_i, T) = \frac{a_c k_{iz}^2}{\pi \phi} n_s \lambda_{HAS} \left\{ A \hbar \omega_0 \coth\left(\frac{\hbar \omega_0}{2k_B T}\right) + (1-A) \hbar \omega_1 \coth\left(\frac{\hbar \omega_1}{2k_B T}\right) \right\}, \quad (1)$$

where $\omega_0 = 1.7$ meV and $\omega_1 = 13.5$ meV are the frequency of moiré phonons and the surface Debye frequency of bulk NbSe$_2$ (taken as the bulk Debye frequency [29] divided by $\sqrt{2}$ [8]), respectively, and the coefficient $0 \leq A \leq 1$ weighs the two contributions. In the prefactor of Eq. (1) r.h. member, $a_c$ is the surface unit cell area, equal to 10.276 Å$^2$ for the semi-infinite bulk NbSe$_2$ [30] and for the few layers, while for single-layer NbSe$_2$ the value $a_c = 10.914$ Å$^2$ is determined from the measured lattice parameter including the dilation induced by charge transfer. Moreover, $k_{iz}$ is the surface normal component of the incident He atom wavevector, $n_s$ the effective number of conducting layers (including multivalley transition multiplicity [15]) contributing to the surface electron-phonon interaction, and $\phi$ the surface workfunction. For the semi-infinite bulk NbSe$_2$, $\phi = 5.9$ eV [31] and $n_s = 2$, as used for 2H-stacked TMDs [13,14,32]. For few-layer NbSe$_2$ it is also $n_s = 2$, but the robust charge transfer from the SiC to NbSe$_2$ lowers the workfunction to that of the latter, $\phi = 4.6$ eV [18]. For the NbSe$_2$ monolayer ($n_s = 1$) on BLG ($n_s = 6$ [15]) a total $n_s = 7$ is used, together with the workfunction of the gBLG/SiC(0001), $\phi = 4.3$ eV [18]. For a given value of $A$, the mass-enhancement factor $\lambda_{HAS}$ is used as a free parameter so as to obtain the best fit of the $-2W(T)$ shape (equal to that of $\ln(I(T)/I_0)$). In this way $\lambda_{HAS}$ is obtained. The high-temperature limit is equivalent to let $\omega_0$ and $\omega_1$ in Eq. (1) tend to zero, which gives

$$2W(\mathbf{k}_i, T) = \frac{2 a_c k_{iz}^2}{\pi \phi} n_s \lambda_{HAS} k_B T \quad, \tag{2}$$

and therefore

$$\lambda_{HAS} = -\frac{\pi \phi}{2 n_s a_c k_{iz}^2} \frac{\partial \ln[I(T)/I_0]}{k_B \partial T} \quad, \tag{3}$$

as reported in [12].

The best linear fits in the high-$T$ limit (Eq. 2) for all the considered samples are shown in Fig. 5 (straight lines) and give $\lambda_{HAS} = 0.76$ for semi-infinite bulk NbSe$_2$, $\lambda_{HAS} = 0.82 \pm 0.06$ and $0.75 \pm 0.06$ for the 2-3ML-NbSe$_2$ data sets 43meV/65º ($k_{iz}^2 = 14.75$ Å$^{-2}$) and 36meV/60º ($k_{iz}^2 = 17.28$ Å$^{-2}$), respectively. For the NbSe$_2$ monolayer on BLG/SiC the fit gives $\lambda_{HAS} = 0.58 \pm 0.04$ or $0.63 \pm 0.04$ when measured with a 43 meV He beam at 60º incidence or with a 22 meV He beam at 52.7º incidence, respectively.

As discussed above in connection with Eq. (1), the $\lambda_{HAS}$ for the two data sets of single-layer NbSe$_2$ actually receives contributions from two distinct phonon spectral regions: moiré phonons and the lattice dynamics of the NbSe$_2$ layer. Figure 7 shows three fits based on Eq. (1) of the 43meV/60° data, one with only the moiré phonons contribution ($A = 1$), one with an equal share of the moiré and the NbSe$_2$ phonons ($A = 0.5$) and the third with only the NbSe$_2$ monolayer phonons ($A = 0$). The two fits including NbSe$_2$ phonons ($A = 0.5, 0$) yield both $\lambda_{HAS} = 0.60 \pm 0.04$, which is consistent with the high-T limit for the 43 meV/60° data. All three fits adequately account quite well for the experimental slope above 150 K, where the high-T linearity is established. However, below this temperature the two fits for $A \neq 0$ diverge from the one including only moiré phonons and from experiment. This suggests that the moiré phonons make a significant contribution to the electron-phonon interaction at low temperatures. In a sense, they serve as acoustic surface modes that exist in bulk NbSe$_2$, but lose their character and localization at small $\Delta K$ in the film due to penetration into the much stiffer substrate. This specific role of the moiré phonons, which preserve their localization also for $Q \to 0$, is consistent with the observation that the moiré phonons are detected by HAS in those regions of the parallel momentum transfer where they intersect the $S_1$ and $S_3$ branches.

The above values of $\lambda_{HAS}$ in few-layer NbSe$_2$ are only slightly larger than the value found for bulk NbSe$_2$, $\lambda = 0.8$. These values are in the range of the theoretical value for the monolayer reported [33], $\lambda = 0.67$, but smaller than the values calculated for two different CDW configurations ($\lambda = 0.84$ and 1.09) [34]. They also agree with the experimental value $\lambda = 0.75$ from electrical transport and optical measurements for NbSe$_2$ monolayer on sapphire (work function 4.5 eV), but strongly disagree for the bilayer value of $\lambda = 0.33$ [4] and with the values for the bulk NbSe$_2$ [35,36], $\lambda = 0.15$. On the other hand, low-temperature photoemission measurements report $\lambda \cong 0.85 \pm 0.15$ for bulk NbSe$_2$ [37].

**II.e) Superconducting critical temperature**

The superconducting critical temperature $T_c$ can be derived from $\lambda_{HAS}$ via the Allen−Dynes-modified McMillan formula [38,39]

$$T_c = 11.5 \tfrac{K}{meV} \langle \omega \rangle_{meV} \exp\left[-\frac{1.04(1+\lambda_{HAS})}{\lambda_{HAS} - \mu^* - 0.62\lambda_{HAS}\mu^*}\right], \quad (4)$$

with the Coulomb pseudopotential $\mu^* = 0.15$ and $\langle \omega \rangle_{meV} = 12.61$ meV taken from [35]. For the bulk NbSe$_2$ ($\lambda_{HAS} = 0.76 \pm 0.03$) it is found $T_c = 4.9 \pm 0.6$ K, which is smaller than the bulk value. Comparable values of $T_c$ are found from the two sets of data for few-layer NbSe$_2$ (6.0 ± 0.7 K from $\lambda_{HAS} = 0.82 \pm 0.06$ and 4.7 ± 0.6 K from $\lambda_{HAS} = 0.75 \pm 0.06$) when the same values of $\mu^*$ and $\langle \omega \rangle_{meV}$ are used for the NbSe$_2$ multilayers. Encapsulated NbSe$_2$ multilayers with $2 \leq n \leq 8$ have shown a uniform decrease of $T_c$ with thickness just in that range, while the $T_c$ of the monolayer encapsulated with hBN drops to 2 K [40].

For monolayer NbSe$_2$ (curve labelled 43meV/60° in Fig. 7), the best fit value $\lambda_{HAS}$ = 0.58 ± 0.04, obtained by including the moiré phonons ($A = 1$), yields, for $\langle \omega \rangle_{meV} = 12.61$ meV, a smaller value of $T_c$, equal to 1.84 ± 0.20 K. This compares well with previous experimental $T_C$ values obtained for single-layer NbSe$_2$/BLG/SiC using STM/STS ($T_c$ = 1.5 K) [41] and transport measurements $T_c$ = 1.5 K [42]. Despite the possible contributions of the moiré phonons, the graphene interfacing has apparently the effect of depressing superconductivity.

**III. Conclusions**

High-resolution HAS diffraction data from single-ayer NbSe$_2$ on BLG/SiC(0001) reveal a (9×9)0°, approximately reducible to a (3√3×3√3)30° superstructure, commensurate with the underlying BLG lattice. Inelastic HAS data provide, besides a set of dispersion curves of acoustic and lower optical phonons, a soft, dispersionless branch of phonons at 1.7 meV, attributed to the interface localized defects distributed with the superstructure period, and thus termed moiré phonons. The electron-phonon coupling for single-layer $NbSe_2$/BLG has been derived from the temperature dependence of the DW exponent, and compared with those of few-layer $NbSe_2$/BLG/SiC(0001) and bulk $NbSe_2$. The low-temperature behavior of the DW exponent suggests an appreciable contribution of the moiré phonons to the electron-phonon coupling, although the intercalation of the BLG between the $NbSe_2$ monolayer and the SiC substrate is seen to attenuate the electron-phonon coupling with respect to the case of a $NbSe_2$ monolayer directly deposited on an inert substrate.

In conclusion, it has been demonstrated that high-resolution HAS diffraction studies can be advantageously extended to long-period conducting surfaces, e.g., those exhibiting long-period superstructures, in order to obtain a detailed structural information. Moreover, the electron-phonon interaction obtained from the analysis of the HAS-DW factor as a function of temperature proves to be, within the experimental uncertainties, quite reliable, also in the case of particularly complex superconducting 2D structures. The growing interest in twisted moiré structures, realized through the assembling of 2D layered materials, should find in HAS a valuable tool for their structural and dynamical characterization.

**Supporting Information**

Supporting informations is available, containing additional experimental details and complementary results.

**Corresponding Author**

* **Prof. Daniel Farías**

Dpto. de Física de la Materia Condensada

Universidad Autónoma de Madrid

28049 Madrid, SPAIN

Phone: +34 91 497 5550    Fax: +34 91 497 3961

daniel.farias@uam.es

**Author Contributions**

The manuscript was written through contributions of all authors. All authors have given approval to the final version of the manuscript.

**Acknowledgements**

This work has been partially supported by the Spanish Ministerio de Ciencia e Innovación under projects PID2019-109525RB-I00 and PID2023-147466OB-C21. D.F. acknowledges financial support from the Spanish Ministry of Economy and Competitiveness, through the "María de Maeztu" Programme for Units of Excellence in R&D (CEX2018-000805-M). M.M.U. acknowledges support by the ERC Starting grant LINKSPM (Grant 758558) and by the Spanish grant no. PID2020-116619GB-C21 funded by MCIN/AEI/10.13039/501100011033.

# Supplementary Information

# Electron-phonon coupling and phonon dynamics in single-layer NbSe$_2$: the role of moiré phonons

Amjad Al Taleb[1], Wen Wan[2], Giorgio Benedek[2,3],
Miguel M. Ugeda[2,4,5] and Daniel Farías[1,6,7]

[1]*Departamento de Física de la Materia Condensada, Universidad Autónoma de Madrid, 28049 Madrid, Spain*
[2] *Donostia International Physics Center, Paseo Manuel de Lardizábal 4, 20018 San Sebastián, Spain.*
[3]*Dipartimento di Scienza dei Materiali, Università di Milano-Bicocca, 20125 Milano, Italy*
[4]*Ikerbasque, Basque Foundation for Science, 48013 Bilbao, Spain.*
[5]*Centro de Física de Materiales, Paseo Manuel de Lardizábal 5, 20018 San Sebastián, Spain.*
[6]*Instituto Nicolás Cabrera, Universidad Autónoma de Madrid, 28049 Madrid, Spain*
[7]*Condensed Matter Physics Center (IFIMAC), Universidad Autónoma de Madrid, 28049 Madrid, Spain*

## Methods

### Sample preparation

Monolayers of NbSe$_2$ were grown on epitaxial bilayer graphene (BLG) on 6H-SiC(0001) by molecular beam epitaxy at a base pressure of ~3·10$^{-10}$ mbar in our home-made UHV-MBE system at the DIPC in San Sebastian (Spain). SiC wafers with resistivities ρ ~ 120 Ω cm were used. Graphitization of the SiC surface was carried out using an automatized cycling mechanism where the sample was ramped between 700 °C and 1350 °C at a continuous ramping speed of ~20°C/s. The SiC crystal was kept for 30 s at 1350 °C for 80 cycles.

Reflective high energy diffraction (RHEED) was used to monitor the layer growth of NbSe$_2$. During the growth, the BLG/SiC substrate was kept at 570°C. High purity Nb (99.99%) and Se (99.999%) were evaporated using an electron beam evaporator and a standard Knudsen cell, respectively. The Nb:Se flux ratio was kept at 1:30, while evaporating the Se led to a pressure of ~4·10$^{-9}$ mbar (Se atmosphere). Samples were prepared using an evaporation time of 30 min to obtain a coverage of ~1-2 ML. To minimize the presence of atomic defects, evaporation of Se was subsequently kept for additional 5 minutes. Atomic Force Microscopy at ambient conditions was routinely used to optimize the morphology of the NbSe$_2$ layers. The samples used for AFM characterization were not further used for STM.

Lastly, in order to transfer the samples from our MBE to Madrid for HAS measurements, they were capped with a ~10 nm film of Se. The capping layer was easily removed in the UHV-HAS system by annealing the sample at ~300°C.

**HAS measurements**

The single-layer NbSe$_2$ sample was mounted in the HAS chamber in Madrid and heated to 300ºC in ultra-high vacuum (UHV) for an hour to desorb the Se capping layer. After cooling down, the crystalline quality of samples was revealed by the observation in HAS of high specular intensity and diffraction features, like the ones shown in Fig. S-1 below.

The samples were characterized by a set of HAS and time-of-flight measurements which were conducted in the Surface Science Laboratory in Universidad Autónoma de Madrid. The experiments have been carried out in two different systems, both having UHV chambers with base pressures in the low 10$^{-10}$ mbar range. The first system is a He-scattering apparatus in which the detector can rotate by 200º in the scattering plane, defined by the beam direction and the normal to the surface, and by +/-15º in the direction perpendicular to the scattering plane. This allows detection of both in-plane and out-of-plane intensities for fixed incident conditions. The He-diffraction measurements reported in the text have been performed using this system. The second chamber is a high-resolution He-scattering machine with a time-of-flight (TOF) arm and a fixed angle of

108° between the incident and outgoing beam. In both HAS machines, the He atom beam is produced by a free expansion of helium gas through a 10 μm nozzle. The incident beam energy can be varied by changing the nozzle temperature. The sample can be heated by electronic bombardment or cooled with liquid nitrogen to 90K or liquid He (down to 40K). The sample can be moved in the X, Y and Z directions, rotated in the X-Y plane and changed the azimuthal angle (φ), which determines the incidence direction with respect to the lattice vector.

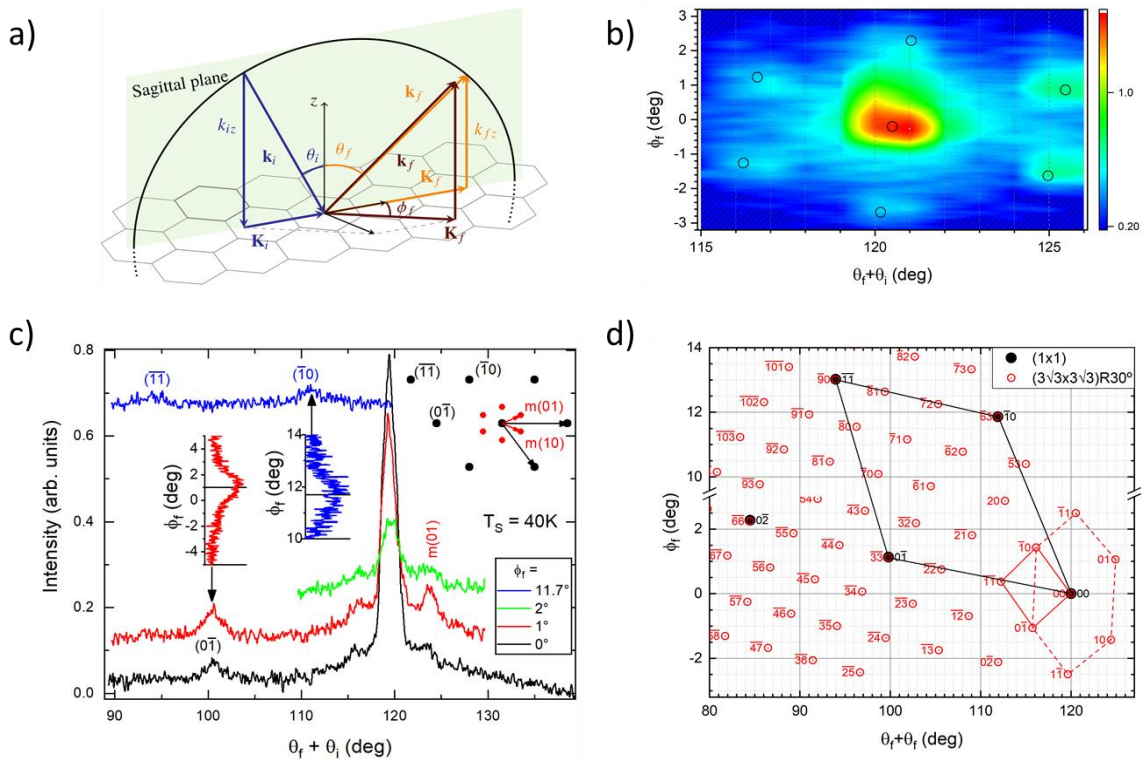

**Figure SI-1.** (a) Kinematics of elastic HAS in-plane and out-of-plane scattering, with $\mathbf{k}_i$ and $\mathbf{k}_f$ the incident and final He-atom wavevectors ($k_i = k_f$), $\theta_i$ and $\theta_f$ the incident and final polar angles, $\phi_i = 0$ and $\phi_f$ the incident and final azimuthal angles with respect to the ΓM direction of NbSe$_2$ (taken as the incident direction). $\mathbf{K}_i$ and $\mathbf{K}_f$ are the respective surface parallel components and $\Delta\mathbf{K} = \mathbf{K}_f - \mathbf{K}_i$. (b) 2D map of the moiré structure overserved on NbSe$_2$ with individual spectra collected performing scans along $\phi_f$. (c) HAS diffraction spectra collected along $\phi_f$ and $\theta_f$ shown as vertical

and horizonal plots, respectively. Note that the maximum of the first order peak ($0\bar{1}$) appears at $\phi_f = 1°$ (red vertical curve), due to azimuthal misalignment of the NbSe$_2$ lattice with respect to the sagittal plane $\phi_i = 0°$. This offset can be traced back to a misalignment $\phi_i = 5°$. **(d)** Calculated diffraction pattern of HAS from both the (1x1) NbSe$_2$ and ($3\sqrt{3}$x$3\sqrt{3}$)R30° surfaces under the experimental conditions: $E_i = 44$ meV, $\phi_i = 5°$ and $\theta_i = 60°$. The measurements shown in b) and c) were done with a sample temperature of 40K.

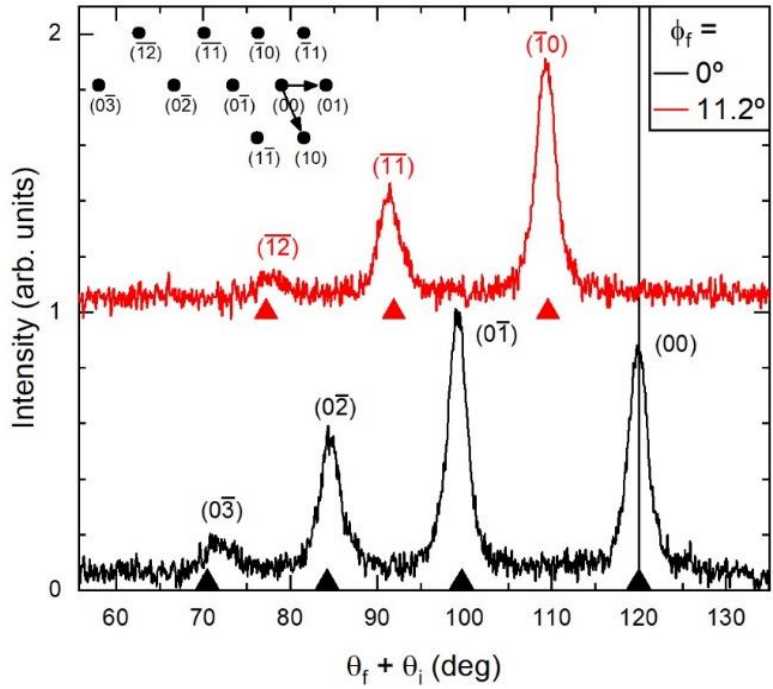

**Figure SI-2.** In-plane (black curve) and out-of-plane (red curve) angular distributions of HAS measured along ΓM from the surface of NbSe$_2$ (0001). The angle of incidence is 60° and $E_i = 44$ meV. Triangles indicate the expected position of diffraction peaks for a lattice constant a=3.44 Å and $\phi_i = 1.5°$ due to azimuthal misalignment of the crystal.